
\font\bigtenrm=cmr10 scaled\magstep2
\magnification 1200
\nopagenumbers
\hsize 14.9truecm \hoffset 1.2truecm
\vsize 22.2truecm\voffset .2truecm
\font \titolo=cmbx12 scaled\magstep0
\outer\def\beginsection#1\par{\medbreak\bigskip
      \message{#1}\leftline{#1}\nobreak\medskip\vskip-\parskip}
\def \TeX{T\kern-.1667em\lower.5ex\hbox{E}\kern-.125em X}
\def \Sg {\Sigma}

\def \om {\omega}

\def \la {\lambda}
\def \La {\Lambda}

\def\de {\delta}
\def \b {\beta}
\def \a {\alpha}

\def \sg {\sigma}
\def \da {\delta}
\def \ep {\epsilon}
\def \part {\partial}

\def \um {{1\over 2}}
\def\one {\rm I}
\def\mk {\bigtenrm {\chi}}
\def\na {a \prime}
\def\nb {b \prime }
\def\nc {c \prime }
\def\det {\rm det}
\def\Tr {\rm Tr}
\def\ch {\rm ch}
\def\sh {\rm sh}
\def\z {\phantom z}

\def\note#1{{\baselineskip=12pt\footnote{$^*$}{#1}}}
%


\def\R{{\rm I\! R}}
\def\\C{{\rm C}}
\def\C{\mkern1mu\raise2.2pt\hbox{$\scriptscriptstyle|$}{\mkern-7mu\rm C}}

\def\and{{\rm\ and\ }}

\def\frac#1#2{{#1 \over #2}}
\def\half{{\frac 12}}
\def\twid #1{\tilde {#1}}

\def \sqr#1#2{{\vcenter{\hrule height.#2pt
       \hbox{\vrule width.#2pt height#1pt \kern#1pt
          \vrule width.#2pt}
       \hrule height.#2pt}}}
\def \square{\mathchoice\sqr68\sqr68\sqr{4.2}6\sqr{3}6}
\def\lsim{\mathrel{\rlap{\lower4pt\hbox{\hskip1pt$\sim$}}
    \raise1pt\hbox{$<$}}}         
\def\gsim{\mathrel{\rlap{\lower4pt\hbox{\hskip1pt$\sim$}}
    \raise1pt\hbox{$>$}}}         

\line{}
\rightline{DFTT 53/93}
\baselineskip=15pt
\leftline{II Workshop on ``Constraints Theory}
\leftline{and Quantisation Methods''}
\leftline{Montepulciano (Siena) 1993}
\vskip 1.0truecm
\centerline{\titolo The Constraints of 2+1 Gravity}
\vskip 0.5truecm
\centerline{J.E.NELSON\footnote{$^*$}{\it email:nelson@to.infn.it}}
\centerline {Dipartimento di Fisica Teorica }
\centerline{Universita degli Studi di Torino}
\centerline {Via Pietro Giuria 1, 10125 Torino, Italy}
\vskip 0.5truecm

The study of the gravitational field in 2+1 spacetime dimensions (2 space,
1 time) has blossomed in the last few years into a substantial industry,
after important contributions by Leutwyler [1], Deser, Jackiw and 'tHooft
[2] and Witten [3]. It provides a means of studying the conceptual
problems it shares with the four dimensional theory. Some of these are
the role of diffeomorphism invariance, time, topology, etc. One of the
advantages over the four dimensional theory is that it is
computationally much easier, although there are many different
interpretations at both the classical and quantum levels.

I shall talk about the constraints of this theory, with emphasis on two
approaches, namely the second order and first order formalisms, and make
comparison with the four dimensional theory wherever possible, and
finally, discuss an operator algebra approach that has been
developed in the last few years in collaboration with T.Regge [4].

The starting point for any study of 2+1 dimensional gravity is the
observation that the Weyl tensor vanishes in 3 dimensions (but not in
4) [5]. It follows that the full Riemann curvature tensor
$R_{\alpha\beta\mu\nu}$
can be decomposed uniquely in terms of only the Ricci tensor
$R_{\mu\nu}$ the scalar curvature $R$ and the metric tensor $g_{\mu\nu}$
itself.
$$
\eqalignno {
	R_{\la\mu\nu k} &=g_{\la\nu} R_{\mu k} - g_{\mu\nu}
	R_{\la k} -g_{\la k} R_{\mu\nu} + g_{\mu k} R_{\la\nu}\cr
	&\quad +\um R(g_{\mu\nu} g_{\la k} - g_{\la\nu} g_{\mu k})  &(1) \cr }
$$

    This is no surprise since the number of independent degrees of freedom
of $R_{\alpha\beta\mu\nu}$ and $R_{\mu\nu}$ are, respectively
${d^2(d^2-1)}\over {12}$ and ${d(d+1)}\over 2$ in $d$ dimensions, and
coincide when $d=3$. If the Einstein tensor $G^{\a\b} = R^{\a\b}
-\um Rg^{\a\b}$ is introduced in (1) the decomposition of
$R_{\la\mu\nu k}$ can then be written
$$ R_{\la\mu\nu k}\ =\ \ep_{\la\mu\b}\ \ep_{\nu k\a}\ G^{\a\b} \eqno
(2)$$
and from (2) it follows that when Einstein's vacuum equations
$$G^{\a\b}= 0 \eqno (3)$$
 are satisfied, the full curvature tensor (all components)
are zero, i.e.$$R_{\la\mu\nu k}\ =0 \eqno (4)$$
and spacetime is flat. In (2) $G^{\a\b}~={{\da S} \over {\da g_{\a\b}}}$
where $S$ is the Einstein-Hilbert action
$$S = \int \sqrt {g}~ R~d^3 x \eqno (5)$$

Thus vacuum solutions of Einstein's equations correspond to flat
spacetimes, and there are no local degrees of freedom. Another, perhaps
more direct, way to see this was given by Carlip [6]. In 3 dimensions,
if the spatial metric tensor $g_{ij},~i,j=1,2$ is the chosen dynamical
variable, its 3 components must satisfy the 3 Hamiltonian constraints,
or equivalently we have the freedom to choose 3 coodinates - 2 spatial
coordinates and time, thus eliminating all local degrees of freedom. In
4 dimensions this is not the case - the spatial metric tensor
$g_{ij},~i,j=1,2,3$ has 6 components subject to 4 constraints, or we may
choose 4 coordinates, leaving 2 degrees of freedom per spacetime point.

To return to flat spacetimes - recall that curvature is defined by
commutators of covariant derivatives, or, by parallel transport around
non-collapsible curves. The change effected by parallel transport around
such closed curves is often called holonomy - and is used to
characterise flat spacetimes. It is here that the topology of spacetime
becomes important - for trivial topologies, for example simply
connected, there are no non-collapsible curves, and equation (4)
follows. There is no dynamics.

It is possible, however, to solve the field equations and introduce some
dynamics, in several ways. The first - developed extensively by Jackiw
et al [2] and others, is to add sources, thus creating local degrees of
freedom. When Einstein's equations read
$$G^{\a\b}= T^{\a\b}$$
where $T^{\a\b}$ is the stress-energy tensor of the sources, the
curvature is no longer zero, but is proportional to $T^{\a\b}$
$$R_{\la\mu\nu k}\ =\ \ep_{\la\mu\b}\ \ep_{\nu k\a}\ T^{\a\b}$$

The simplest example is perhaps that of a point particle [7] at a
spatial point $x$ which would create a spatial curvature there
proportional to its mass
$$^2R(x) = m \da^2 (x) \eqno (6)$$
but even this very simple example has a number of different
interpretations. Firstly, removing a single point from a flat
2-dimensional manifold produces a singularity in the curvature at that
point of the type (6), so it may be useful to study punctured surfaces.
Clearly related to this is the observation that if a 3-dimensional knot
falls through a 2-dimensional surface (or a dynamical 2-surface
passes through a 3-dimensional knot), the knot will puncture the surface
in an even number of points. These punctures will then move around the
surface, and their number will change. This can be thought of as the
dynamics of point particles, and their annihilation and creation [8].
Another interpretation is provided by Regge calculus [9]. If a curved 2-
surface is approximated by a number of flat triangles (for example),
then the curvature of the surface is only non-zero at the vertices
$z_i$ say, of the triangles, and is proportional to [10]
$$
\z ^2 R(z) \propto  {\sum _{z_i}} \, \da^2 (z-z_i)
$$
analogous to (6).

Propagating massive gravitational modes can be generated by adding a
topological term to the action (5), always possible in an odd number of
dimensions [5]. For gravity in 3 dimensions, this is the Chern-Simons
form
$$\int (\om^{ab}\wedge d\om_{ab}+{2\over 3}\om^{ac}\wedge\om^
{d}_{c}\wedge\om_{da}) \eqno (7)$$
where the components of the spin connection $\om^{ab}_{\mu}$ are to
be considered as
functionals of the triads $e^{a}_{\mu}$ by solving the torsion equation.
$$R^a = de^{a}-\om^{ab}\wedge e_{b}\ = 0 $$
with $e^{^a}_{\mu}e^{^b}_{\nu}\eta_{ab}=g_{\mu\nu}$.
Variation of (7) with respect to the metric tensor $g_{\mu\nu}$ gives
the Cotton tensor
$$C^{\mu\nu}=g^{-\um}\ \ep^{\mu\la\b}D_{\la}\left( R^{\nu}_{\b}-{1\over
4} \da^{\nu}_{\b}R\right) $$
which is symmetric, traceless, conserved, and vanishes if the theory is
conformally invariant. Therefore, adding the Chern-Simons term (7) to
the scalar curvature action (5) with a constant factor ${1\over \mu}$
leads to the field equations
$$G^{\mu\nu}+{1\over \mu}C^{\mu\nu}=0$$
which can be transformed into
$$\Big(\square +\mu^{2}\Big)\ R_{\mu\nu}=\ {\rm terms\ in}\ \Big(
R_{\mu\nu}\Big)^{2} \eqno (8)$$

In the linearized limit the R.H.S. of (8) vanishes and it is shown in
[5] that the solutions of (8) correspond to massive, spin $\pm 2$,
particles.

A way to introduce global but non-local degrees of freedom is to
consider topologies with non-collapsible curves. A simple example is
when the spatial surfaces are tori - then the meridian and parallel are
clearly non-collapsible (or, non-homotopic to the identity). I shall
talk about this in more detail later.

I'd like now to discuss the constraints of 2+1 gravity for the Einstein-
Hilbert action (5) in the canonical formalism, where spacetime is
$R\times \Sg$, and time runs along $R$. There are many ways to write
them and many ways to interpret them. There is firstly the choice of
canonical variables but these are largely equivalent. However there are
differences between the first and second order formalisms.

In the second order formalism take as variables the spatial metric
tensor $g_{ij},~i,j=1,2$ and its conjugate momentum
$\pi^{ij}= {\sqrt {g}}(K^{ij}-Kg^{ij})$ where $K^{ij}$  is the extrinsic
curvature. The action (5) is decomposed as
$$\int(\pi^{ij} {\dot g}_{ij}- N^i{\cal H}_i - N{\cal H})d^3x$$
The lapse $N^i$ and shift $N$ functions are related to the non-dynamical
components of $g_{ij}$ and their variation leads to the constraints on
$g_{ij}$ ,$\pi^{kl}$
$$\eqalignno{
{\cal H}^{i}\ &=\ D_{j}\ p^{ij} \approx 0 \qquad\qquad\qquad i,j=1,2.
& (9) \cr
{\cal H}_{\perp}\ &=\ p_{ij}\ p^{ij}-\ \Big(p^{ij}\ g_{ij}\Big)^{2}-
^{2}R \approx 0 & (10) \cr} $$
where the covariant derivative $D_{j}$ is with respect to $g_{ij}$. The
constraints (9), (10) are formally the same as in 4 dimensions [11], are
non-polynomial in $g_{ij}$ ,$\pi^{kl}$ and involve $(g_{ij})^{-1}$. With
the Poisson brackets
$$[g_{ij}(x),\pi^{kl}(y)]= \um (\da^{k}_{i}\da^{l}_{j}+\da^{k}_{j}\da^{l}
_{i})\da^{2}(x-y) \eqno (11)$$
the constraints ${\cal H}^i$ generate, as in 4 dimensions,
diffeomorphisms (or coordinate transformations) in the spatial surface
$\Sg$ through the bracket (11). But the generator of dynamics, or time
reparametrisation invariance, is the full Hamiltonian, namely the
combination
$$\int( N^i{\cal H}_i + N{\cal H})d^3x$$

In the first order formalism the situation is rather different. Here the
conjugate variables are the spatial triads $ e^{a}_{i}$
related to the spatial metric by $e^{a}_{i}e^{b}_{j}\eta_{ab}=g_{ij}$,
and the components $ \ep^{ij} \ep _{abc} \om^{{ab}}_{i}$
of the spin connection. The action (5)  decomposes as
$$\int (\ep^{ij} \ep _{abc} \om^{{bc}}_{j} {\dot e}^{a}_{i}
- e^a_0{\cal H}_a - \om^{ab}_0 J_{ab})d^3x \eqno (12)$$
where
$${\cal H}_{a} = \ep^{ij}\ \ep_{{abc}}\ R_{ij}^{{bc}} \eqno (13)$$
$$J_{{ab}} = \ep^{ij}\ \ep_{{abc}}\ R_{ij}^{c}  \eqno (14)$$
with $a,b,c =0,1,2, i,j = 1,2, \ep_{ij}=-\ep_{ij}, \ep_{12}=1$
and $R_{ij}^{{bc}}$, $R_{ij}^{c}$ are the spatial components of the
curvature
$$R_{ij}^{{ab}}=\omega^{{ab}}_{j,i} -\omega^{{ac}}_i
\omega_{cj}^{{b}}-(i\leftrightarrow j) \eqno(15)$$
and torsion
$$R_{ij}^{a} = e^{a}_{j,i} -\om^{{ab}}_i e_{b j}
-(i\leftrightarrow j)
\eqno (16) $$
respectively. Since $e^a_0$, $\om^{ab}_0$ are non-dynamical, the
variational equations (constraints) derived from (12) are evidently
$${\cal H}_{a} \approx 0$$
$$J_{{ab}} \approx 0 $$
or alternatively
$$R_{ij}^{{bc}} \approx 0 \eqno (17)$$
$$R_{ij}^{c} \approx 0 \eqno (18)$$

With the Poisson brackets
\note{ This is in marked contrast to the 4-dimensional case
[13] where the constraints read
$$
\eqalignno { &R _{[ij}^{bc} \, e ^a_ {k]}
\, {\ep} _{abcd} = 0 \qquad i,j,k = 1 , 2 , 3            &(20)\cr
&R _{[ij]}^a = 0 \qquad a , b , c , d , = 0 , 1, 2, 3
                                                    &(21)\cr}
$$
and the conjugate variables are $\om^{{ab}}_{i}$ and the combination
$(det e) \ep^{ijk} \ep _{abcd} e^{c}_{j} e^{d}_{k}$
so it is not possible to write Poisson brackets  between the
$\om^{{ab}}_{i}$ and the  $e^{a}_{i}$ directly. It is only in the time
gauge $e^{0} _{i} = 0$ that the remaining dyads $e ^{\na}_i$, ${\na} =1, 2, 3$
satisfy the identity
$$
\eqalignno { e ^{\na} _i e ^{\nb} _j e ^{\nc} _k \ep _{\na \nb \nc} & =
\ep _{ijk}                                  \cr}
$$
and therefore the conjugate to $\om _i ^{\na \nb}$ is zero whereas
the conjugate to $\om _i ^{ 0 \na }$ is the contravariant density
$(\det e) e ^i _{ \na }$. These are the variables used in the new
variables formulation of non-perturbative gravity [14].}
$$
[e^{{a}}_{i}(x),\om^{{bc}}_{j}(y)]
=-\ep_{ij}\ep^{abc}\da^{2} (x-y) \eqno (19)
$$
where $x,y~ \ep~\Sg$ the constraints (17), (18) satisfy the $ISO(2,1)$
or Poincare' algebra in 3 dimensions, that is, involving structure
constants only (in 3+1 dimensions there are structure functions which
involve the components of the curvature [12]). The torsion components
(18) generate $SO(2,1)$ rotations ($\da e^{a}_i = \la^{a}_b e^b _i,
\da \om^{{ab}}_i = D_i\la^{ab}$) for parameter $\la^{ab}$) whereas
the curvature components (17) generate diffeomorphisms, or coordinate
transformations ($\da e^{a}_i = D_i u^{a}, \da \om^{{ab}}_i =0$) for
parameter $u^{a}$) in $\Sg$.

Having introduced the constraints (9,10) and (17,18) of 2+1 gravity it
is interesting to see what we could expect if they were quantised and
then applied as operators on wave functions $\Psi(g)$ or $\Psi(e,\om)$.
In the second order (metric) formalism, if the momenta $\pi^{ij}$ are
applied as operators of the form
$$\pi^{ij}(x) \sim  {\da \over {\da g_{ij}(x)}}$$
then ${\cal H}_i \Psi(g) =0$ with ${\cal H}_i$ given by (9) implies
that one should identify wave functions of metrics $g_{ij}$ and
$\twid {g} _{ij}$ when they differ as
$$
\twid {g} _{ij} =g_{ij} +D_i N_j+D_j N_i
$$

This is just the Lie derivative in the direction $N^i$, or the result of
a coordinate transformation in $\Sg$. The same analysis holds in 4
dimensions. The constraint ${\cal H} \Psi(g) =0$ with ${\cal H}$ given
by (10) is much more difficult to interpret, as in 4 dimensions,
although some partial progress has been made recently [6].

Things are a little clearer in the first order, triad, spin connection,
formalism. Here we must use the connections $\om^{{ab}}_{i}$ as
coordinates and treat the triads $e^{a}_{i}$ as momenta [3] in order to
have diffeomorphism invariant wave functions ($\da e^{a}_i = D_i u^{a}$
depends on the connection whereas $\da \om^{{ab}}_i =0$). So if we assign
$$
e^{a}_i(x) \sim \ep_{ij} \ep ^{abc} {\da \over {\da \om^{bc}_{j}(x)}}
$$
and $\om^{{ab}}_i$ acting multiplicatively, then from (15)
$$
\eqalignno { R_{ij} ^{ab} ( \om ) \, \Psi ( \om ) &= 0 \cr}
$$
means that $\Psi ( \om )$ is defined only on flat connections.
There is no such interpretation in 4 dimensions - in fact the curvature
there is non-zero, c.f.(20). However, the torsion constraint (16)
$$
\eqalignno { R _{ij} ^a ( e , \om ) \, \Psi ( \om ) &= 0 \cr}
$$
is the same in both 3 and 4 dimensions and implies here that we should
identify wave functions $\Psi ( \om )$ for connections $\om$
and $\twid {\om}$ that differ by an $SO(2,1)$ rotation
$$
\eqalignno { \twid {\om} _i^{ab} &= \om _i^{ab} +
D _i \la ^{ab}                                  \cr}
$$
that is,
$$
\eqalignno { \Psi ( \om ) &= \Psi ( \twid {\om} )
                                                \cr}
$$
and is $SO(2,1)$ invariant.

    One remarkable difference between 3 and 4 dimensions is that in 3
dimensions the constraints can be solved classically in both the second
and first order formalisms.

In the second order, metric formalism, one can always write the metric
$g _{ij}$ of the (now compact) spatial Riemann surface $\Sg$ of genus
g in the form
$$
\eqalignno { g _{ij} &= e ^{2\la} \twid {g} _{ij}       \cr}
$$
where ${g _{ij}}$ is a constant curvature metric on $\Sg$ with
$$
\eqalignno { \z ^2 R ( \twid {g} ) = k \, , \quad &\quad
\matrix {k = 1 \quad & \quad g = 0       \cr
         k = 0 \quad & \quad g = 1       \cr
        k = -1 \quad & \quad g > 1 \, .  \cr}       \cr}
$$

    For the particular choice of time (gauge)
$T=Tr K= - {g}^{-\um}g_{ij} \pi^{ij}$ the Hamiltonian constraint (10)
reduces to a  differential equation for the conformal factor $\la$
[15]
$$
\eqalignno { {\cal H}_{\perp} &= \Delta _{\twid {g}} \la -
{T ^2\over 4} e ^{2\la}
+ {\half} \twid {g} ^{-1} \twid {\pi} _{ij} \twid {\pi} ^{ij} \,
e ^{-2\la} - {k\over 2} = 0                     & (22) \cr}
$$
where ${\twid \pi}^{ij}$ is the traceless transverse part of the
momentum conjugate to $\twid {g} _{ij}$, and (9) is automatically
satisfied. A solution of (22) for $\la$ always exists for $g \ge 1$, and
the action (5) becomes
$$
\eqalignno { S &= \int \! d T \Bigl ( p ^{\a}
{d {\tau} _{\a}\over d T} - H ( p , {\tau} ) \Bigr )
                                \, \, \, .        & (23) \cr}
$$
where $\tau_{\a}$ are coordinates on Teichmuller space, $\a=1....6g-6$
and have conjugate momenta
$$
\eqalignno { p ^{\a} &= \int _{\Sigma} \! d ^2 x
{\twid \pi}^{ij} {\partial \over \partial \tau _{\a}}
\twid {g} _{ij}                                 \cr}
$$
In (23) $H(p,\tau)$ is an effective Hamiltonian representing the area
of $\Sg$ at time $T$
$$
\eqalignno { H &= \int _{\Sigma} \! d ^2  x \, g ^{\half}
= \int _{\Sigma} \! d ^2 x \, {\twid {g}} ^{\half} e ^{2\la} \cr}
$$
where $\la=\la(\twid g,\twid \pi,T)$ is given by (22). So $H(p,\tau)$
generates time development in the gauge $T=Tr K$. For $k=0$, $g=1$, the
torus, (22) is easily solved and $\tau= \tau_1 +i\tau_2$ are the two
degrees of freedom (the moduli) of the flat metric $\twid {g} _{ij}$ of
$\Sg$. This programme is extremely interesting, although for $g~>~1$ it
seems much harder to solve (22). The role of the gauge $T=Tr K$ has
yet to be clarified.

    The constraints of 2+1 gravity can also be solved in the first order
formalism, and this is where I have been most involved [16,17]. To
recall, these constraints (17,18) read
$$
\eqalignno { R ^{ab} &= d \om ^{ab} - \om ^{ac} \land {\om _c} ^b = 0 \cr
R ^a &= d e ^a - \om ^{ac} \land e _b = 0    \cr}
$$
restricted to $\Sg$. They can be solved locally by setting
$$
\eqalignno { \om ^{ab} &= d G ^{ac} {G _c} ^b                     \cr
                e ^a  &= d J ^a - d G ^{ac} {G _c} ^b J _b       \cr
                      &= d J ^a - \om ^{ab}    J _b              \cr}
$$
where $G ^{ab} \in SO (2 , 1)$ and $J ^a \in \R ^3$
together form an element $\mk$ of the Poincare group $ISO(2,1)$ in 3
dimensions represented by the $4\times 4$ matrix
$$
\eqalignno { \mk  &=  \left ( \matrix {
G ^{ab} & J ^b \cr 0 & 1 \cr } \right )
                                               &(24) \cr}
$$
with
$$
\eqalignno { d \mk \, \mk ^{-1} &= \left ( \matrix {
\om & e \cr 0 & 0 \cr } \right )
                                                    \cr}
$$
Matrices of the type (24) can be used to provide a representation
$\mk (\pi _1)$ of the fundamental group of the Riemann surface
$\Sg$ as follows.

    If the fundamental group $\pi _1$ of the surface $\Sg$ has
generators $U_{1},V_{1},.....U_{g},V_{g}$, for genus g,
assign to each generator a matrix of the form (24)
$$
\eqalignno { \mk (U _k)  =  \left ( \matrix {
G _k & J _k \cr 0 & 1 \cr } \right )
\quad & \quad
\mk (V _k) =  \left ( \matrix {
F _k & K _k \cr 0 & 1 \cr } \right )
                                               \cr}
$$
satisfying
$$
\eqalignno { \mk (U \, V)  &=  \left ( \matrix {
G F & J + G K \cr 0 & 1 \cr } \right )
                                                \cr
\mk (U ^{-1}) &=  \left ( \matrix {
G ^{-1} & -G ^{-1} J  \cr 0 & 1 \cr } \right )
                                               \cr}
$$

    The Poisson brackets between the matrices $\mk$
can be obtained by integrating the brackets (19) along infinitesimal
intersecting paths on $\Sg$, and then assembling to form finite closed
paths [16]. But the relations obtained depend on the base point of the
paths (clearly the base points can be moved around $\Sg$ by
diffeomorphisms). One way to construct invariant quantities is to take
traces since for example,
$$
\eqalignno { \Tr \, G ^{ab} ( \rho ) &= G ( \rho )
= G ( \da ^{-1} \rho \da )                              \cr}
$$
if $\rho$ is a closed path and $\da$ any open path on $\Sg$. Another
invariant quantity is defined by
$$
\eqalignno { \nu ( \sg ) &= G ^{ab} ( \sg ) J ^c ( \sg )
\ep _{abc} = \nu ( \da ^{-1} \sg \da )              \cr}
$$
These quantities satisfy the Poisson brackets
$$
\eqalignno { [ G ( \rho ) \, , \, \nu (\sg) ] &= G ( \rho \sg )
- G ( \rho \sg ^{-1} )                                      \cr
 [ \nu ( \rho ) \, , \, \nu (\sg) ] &= \nu ( \rho \sg )
- \nu ( \rho \sg ^{-1} )                     \, \, \, . & (25)\cr}
$$
valid when the paths $\rho, \sg$ have a single intersection. The algebra
(25) is very similar to the loop algebras of the new variables 3+1
gravity [14].

The algebra (25) and its representations are most easily studied by
considering the spinor group $SL(2,\R)$ with elements $S^{\a}_{\b}$
related to the $SO(2,1)$ elements $G_{ab}$ through
$$
S \tau_b S^{-1} =\tau^a G_{ab}
$$
where the $\tau_a$ are the pseudo Pauli matrices
$$
\tau _a \tau _b = \eta _{ab} + \ep _{abc} \tau ^c \, \, \, .
$$

    It is also convenient to introduce a cosmological constant $\La$ [17] to
the action (5). Define an extended connection $\om ^{AB}$ which includes
all components of $ e,\om$
$$
\eqalignno { {\om ^A} _B  &=  \left ( \matrix {
{\om ^a} _b & {s\over \a} e ^a \cr
- {s\over \a} e _b  & 0 \cr } \right )
                                               \cr}
$$
where $A , B = 0 , 1, 2, 3$; $a, b = 0 , 1, 2$; $s = \pm 1$;
$\La = {s\over \a ^2}$ and indices are raised and lowered with
the (flat) metric $\eta_{AB}=(-1,1,1,s)$. So the gauge group
now is $SO(3,1)$ or $SO(2,2)$ with corresponding spinor groups
$SL (2, \\C )$ or $SL (2, \R ) \otimes SL (2, \R )$
respectively, and $\om^{AB}$ is to be identified with
$\om^{AB}=(dE E^{-1})^{AB}$ where $E ^{AB} \in SO(3,1)$ or $SO (2,2)$
and $E ^{AB} \to   \left ( \matrix { G ^{ab} & J ^a \cr
0  & 1 \cr } \right )$ as $\La \to 0$.

    In the spinor representation with $\La < 0$, $\La= -{1\over \a ^2}$
$$
\eqalignno { E ^{AB} \gamma _B &= S ^{-1} \gamma ^A S \, ,
\quad  S \in SL (2, \R ) \otimes SL (2, \R )        \cr}
$$
with Dirac matrices $\gamma ^A$ satisfying $\{ \gamma _A \, ,
\, \gamma _B \} = 2 \eta _{AB}$ the brackets (25) are united in the
single Poisson bracket
$$
\eqalignno { [ S ( \rho ) \, , \, S (\sg) ] &= {1\over 4\a}
\Bigl ( S ( \rho \sg ) -  S (\rho \sg ^{-1} )  \Bigr )   &(26) \cr}
$$
In (26) $S ( \rho )= \um Tr S^{\a}_{\b}( \rho)= \um S^{\a}_{\a}( \rho)$
but since there is the identity
$$
\eqalignno { \Tr (A \, B) + \Tr (A \, B^{-1} ) &=
\Tr(A) \, \Tr(B)                                \cr}
$$
between $2\times 2$ matrices A,B, (26) becomes
$$
\eqalignno { [ S ( \rho ) \, , \, S (\de) ] &= {1\over 2\a}
\Bigl ( S ( \rho \de ) - S (\rho) \,  S (\de ) \Bigr)
                                            &(27) \cr}
$$
It is this version (27) of the algebra of observables that has been most
studied [17].

    I shall now discuss the case g=1 $\La < 0$ in more detail.
Since for the torus
there are only 3 independent paths $U,V,UV$, all with single
intersections, satisfying
$$
\eqalignno { U \, V \, U ^{-1} \, V ^{-1} & = \one             \cr}
$$
we can form 3 traces
$$
\eqalignno { x = \Tr \Bigl ( S (U) \Bigr )   \quad
             y &= \Tr \Bigl ( S (V) \Bigr )  \quad
             z = \Tr \Bigl ( S (U \, V) \Bigr ) &(28) \cr}
$$
which must satisfy
$$
\eqalignno { F &= \Tr \Bigl ( S (U \, V \, U ^{-1} \, V ^{-1})
- \one \Bigr )                                      \cr
               &= 1 - x ^2 - y ^2 - z ^2 + 2 xyz = 0 \, . &(29)\cr}
$$

    The expression (29) has zero brackets with all $x,y,z$ and is
symmetric under cyclical permutations of $x,y,z$. The brackets (27) are
then
$$
\eqalignno { [ x \, , \, y ] &= {1\over 2\a} ( z - x \, y )  &(30)\cr}
$$
and are also cyclically symmetric. Note that $F=0$ can be solved by
setting $x= \cos a$, $y= \cos b$, $z= \cos (a+b)$, and then (30)
implies, for the two remaining variables $a$, $b$
$$ [ a \, , \, b ] =- {1\over 2\a} $$

The algebra (30) is highly symmetrical. Since, for example,
$$
\eqalignno { [ x \, , \, y ] &= {1\over 2\a} ( z - x \, y )        \cr
             [ x \, , \, z ] &= {1\over 2\a} ( x \, z - y )        \cr}
$$
one can think of $x$ as generating a canonical transformation on the
other two variables $y,z$, but mixing them. If $y(0)$ and $z(0)$
represent the untransformed variables then the equation
$$
\eqalignno { [ x \, , \, y (t) + \la z (t) ] &=
\mu \, ( y (t) + \la z (t) )                   \cr
&= {d\over dt} (y (t) + \la z (t) )             &(31)\cr}
$$
can be solved (with parameter $t$) for eigenvalues
$\mu = \mp 2 \sh \phi$, and eigenvectors
$y (t) +\la z (t)$ with $\la = - e^{\pm \phi}$
and $x= \ch \phi$. The result is [18]
$$
\eqalignno { \left ( \matrix { y(t) \cr
                       z(t) \cr } \right ) &=
{1\over \sh \phi} \,
\left ( \matrix { {\sh \phi (1+t)\over 2} & - \sh \phi t \cr
          \sh \phi t     & {\sh \phi (1-t)\over 2} \cr } \right )
\, \left ( \matrix { y(0) \cr z(0) \cr } \right )           \cr
&= \Omega \, \left ( \matrix { y(0) \cr z(0) \cr } \right )
                                                        &(32)\cr}
$$
so one can think of $x$ as acting like a Hamiltonian for $y(t)$ and
$z(t)$ and generating their development in the parameter $t$. If $t$ is
integral then the transformation (32) is polynomial and
$\Omega (n) = \Omega (1) ^n$ so we may take $t=1$.
The transformation $D_x$ generated by $x$ is then
$$
\eqalignno { D_x : y &\to x y - z            \cr
                   z &\to y                  \cr
                   x &\to x            &(33) \cr}
$$
and comparing with the traces (28) this corresponds to
$H _U : U \to U$, $V \to V \, U^{-1}$ which is a Dehn twist of the torus.
Each element $x,y,z$ generates a transformation $D_x,D_y,D_z$ satisfying
$$
\eqalignno { D_y \, D_x \, D _y &= D_x \, D_y \, D _x
\qquad and \quad cyclical.                           &(34)\cr}
$$
Equation (34) is just one of the Braid group identities for 3 cyclical
strings. Similarly for the fundamental group we have
$$
\eqalignno { H _U \, H _V \, H _U &= H _V \, H _U \, H _V \cr}
$$
that is, one of the identities of the Dehn or mapping class group.

It has been shown [19] that for g=1 and $\La=0$ the second order
(metric) and first order (connection) formulations are classically
equivalent. They are related through a time dependent canonical
transformation. There are similar, recent results for $\La \ne 0$
[20]. In the quantum case this is very unclear and I now turn
to the quantisation of the connection formalism.

The classical algebra (30) and its symmetries (34) are quantised as
follows. In (30)  replace the bracket $[ x , y ]$ by
${xy-yx\over i \hbar}$ and on the R.H.S. , $xy$ by
${\half} \, (xy +yx)$. The result is
$$
\eqalignno { e ^{i \theta} xy - e ^{-i \theta} yx &=
2i \sin \theta z
\qquad and \quad cyclical               &(35)\cr}
$$
with $\tan \theta =  {\hbar \over 4\a}$.

The expression (35) is the cyclical representation of $SU(2)_q$ [17]
with parameter $q = e ^{2i \theta}$, and its q-Casimir is
$$
\eqalignno { F &= 2e^{i\um \theta } xyz - e^{-i\theta } (x^{2}+z^{2})
-e^{- i\theta } y^{2} \, , \qquad cyclically \quad invariant
                                        \, \, \,          &(36) \cr}
$$
The transformations (Dehn twists) (33) are replaced by
$$
\eqalignno { y &\to {(1+K)\over 2} \, x y - z            \cr
             z &\to y                               &(37)\cr}
$$
with $K = e ^{i\theta}$. These quantum, ordered, transformations are
generated by [4]
$$
\eqalignno { y &\to F (\psi) y F (\psi) ^{-1}           \cr
             z &\to F (\psi) z F (\psi) ^{-1}           \cr}
$$
with
$F (\psi) = \exp ^{\bigl ( -i {\psi ^2 \over 2\theta} \bigr )}$,
$x = {\cos \psi \over \cos {\theta\over 2}}$.

    The quantum algebra (35) and quantum Casimir (36) are invariant
under (37). For g=1 there is a Heisenberg-Weyl representation [4] of
(35) which splits into subrepresentations of dimension m, for some m,
each acting like the identity, when ${\theta \over \pi}$ is rational.
This happens when $q = e ^{2i\theta }$ is the root of unity.

    The algebra (35) has been extended to arbitrary genus [4], when
there are paths in $\Sg$ which have more than one intersection. It is
an abstract quantum algebra invariant under the quantum Dehn twists
(37). For arbitrary g the number of independent variables is 6g-6,
the number of independent moduli [21].

    Some unanswered questions in this last approach are for example,
the role of time and time development, and the relationship with the
metric formalism [15] for $g>1$. It would seem that there is some
connection with the four dimensional theory but this has yet to be
established.
\vskip 15truept
\centerline {\bf References}

\vskip .7truecm
\item{[1]\ }H.~Leutwyler, il Nuovo Cimento Vol.{\bf XLIIA} (1966)159.
\item{[2]\ }S.~Deser, R.~Jackiw and G.~'tHooft, Ann. Phys.{\bf 152}
(1984)220.
\item{[3]\ }E.~Witten, Nucl.Phys.{\bf B311} (1988/89)46-78.
\item{[4]\ }J.~E.~Nelson and T.~Regge, Phys.Lett.{\bf B272} (1991)213
and references therein.
\item{[5]\ }S.~Deser, R.~Jackiw and S.~Templeton, Ann.Phys.{\bf 140}
(1982)372.
\item{[6]\ }S.~Carlip, in proceedings of Fifth Canadian Conference on
General Relativity and Relativistic Astrophysics, Ontario, May 1993.
\item{[7]\ }R.~Jackiw, in proceedings of XVII International Colloquium
on Group Theoretical Methods in Physics, Canada, June 1988.
\item{[8]\ }L.~Kaufmann, private discussion.
\item{[9]\ }T.~Regge, il Nuovo Cimento, {\bf 19} (1961)558.
\item{[10]\ }D.~Foerster, Nucl.Phys.{\bf B283} (1987)669.
\item{[11]\ }R.~Arnowitt, S.~Deser and C.~Misner, in ``Gravitation:An
Introduction to Current Research'' (L.Witten,ed.) Wiley, NY 1962.
\item{[12]\ }J.~E.~Nelson and C.~Teitelboim, Ann.Phys.{\bf 116} (1978)1.
\item{[13]\ }J.~E.~Nelson and T.~Regge, Ann.Phys.{\bf 166} (1986)234.
\item{[14]\ }A.~Ashtekar, ``Lectures on Non-Perturbative Canonical
Gravity'', World Scientific, Singapore (1991).
\item{[15]\ }V.~Moncrief, J.Math.Phys.{\bf 30} (1989)2907.
\item{[16]\ }J.~E.~Nelson and T.Regge, Nucl.Phys. {\bf B328} (1989)
190.
\item{[17]\ }J.~E.~Nelson, T.~Regge and F.~Zertuche, Nucl.Phys.
{\bf B339} (1990)516.
\item{[18]\ }J.~E.~Nelson and T.Regge, C.M.P.{\bf 141} (1991)211.
\item{[19]\ }S.~Carlip, Phys.Rev {\bf D42} (1990)2647.
\item{[20]\ }S.~Carlip and J.~E.~Nelson, in preparation.
\item{[21]\ }J.~E.~Nelson and T.Regge, C.M.P.{\bf 155} (1993)561.

\vfill\eject\end